\documentclass[preprint,prd,aps,showpacs,showkeys,nofootinbib]{revtex4}
\usepackage{graphicx}
\begin{document}


\title{The four-propagator three-loop vacuum integral by the hypergeometry }

\author{ Zhi-Hua Gu$^{a,b}$\footnote{email:sxguzhihua@hebau.edu.cn},
 Hai-Bin Zhang$^a$\footnote{email:hbzhang@hbu.edu.cn}}

\affiliation{$^a$Department of Physics, Hebei University, Baoding, 071002, China}
\affiliation{$^b$College of science, Agricultural University of Hebei, Baoding, 071001, China}

\begin{abstract}
Hypergeometric function method is proposed to calculate the scalar integrals of Feynman diagrams.
For the scalar integral of three-loop vacuum diagram with four-propagator, we verify the equivalency of Feynman parametrization and
the hypergeometric technique. The result can be described as generalized hypergeometric functions of triple variables.
 Based on the triple hypergeometric functions, we also establish the systems
of homogeneous linear partial differential equations(PDEs) satisfied by the mentioned scalar integral.
The continuation of the scalar integral from its convergent regions to whole kinematic domains can be made numerically
through the system of homogeneous linear PDEs with the help of the element method.
\end{abstract}


\maketitle

\section{Introduction\label{sec1}}
\indent\indent
Applying the method of integration by part (IBP)\cite{K.G.Chetyrkin81},
the general scalar integrals can be reduced to a linear combination of scalar integrals.
Calculating scalar integral is an obstacle to predict the electroweak observables precisely
in the standard model. Those one-loop scalar integrals are computed\cite{G.'t.Hooft79,A.Denner11},
however the calculations of the multi-loop scalar integrals
are not advanced enough. A number of useful methods are introduced
to evaluate those scalar integrals in literature\cite{V.A.Smirnov12}.
An analytic expression for the planar massless two-loop vertex diagrams is given in Ref.\cite{R.J.Gonsalves83} by Feynman parametrization method.
The Mellin-Barnes (MB) method is sometimes used to compute some massless scalar integrals\cite{V.A.Smirnov99,J.B.Tausk99}. In the paper\cite{E.E.Boos}, the Mellin-Barnes representation is used to obtain expressions for some classes of single-loop massive Feynman integrals and vertex type.
The results are presented in the form of hypergeometric functions.
Furthermore, multivariable hypergeometric functions are presented giving explicit
series for small and large momenta for two-loop self-energy diagrams\cite{S.Bauberger}.
However, the technique of multiple MB representations will be very cumbersome for multi-loop diagram.
For scalar integral, the author of Refs.\cite{A.V.Kotikov91-1,A.V.Kotikov91-2,A.V.Kotikov91-3,A.V.Kotikov91-4,
A.V.Kotikov92-1,S.Laporta96,S.Laporta97,E.Remiddi97,S.Laporta00,K.Melnikov00-1,K.Melnikov00-2,
M.Y.Kalmykov10,M.Y.Kalmykov12,M.Y.Kalmykov13,M.Y.Kalmykov17} derives a set of differential equations based on the IBP relationship.
Another method is recommanded to analyze the scalar integrals which is called ``dimensional recurrence
and analyticity'' \cite{R.N.Lee10-1,R.N.Lee10-2,R.N.Lee11-1,R.N.Lee11-2,
R.N.Lee11-3,R.N.Lee12-1,R.N.Lee12-2}.
The asymptotic expansions in momenta and masses can be employed to approach the scalar integral relies on kinematic invariants
and masses\cite{V.A.Smirnov02}.
A novel method \cite{Xiao liu1,Xiao liu2} to compute
Feynman  integrals by constructing and solving a system of
ordinary differential equations (ODEs).\\
\indent\indent
The class of two-loop massive scalar
self-energy diagrams with three propagators is studied in
an arbitrary number of dimensions\cite{F.A.Berends}, and they can be described by
generalized hypergeometric functions of several variables,
namely Laricella functions. The results can be generalized to N loop massive scalar
self-energy diagrams with N + 1 propagators.
But they only get the analytical results in the converges.
The continuation  from its convergent regions
to whole kinematic domain has not been finished.\\
\indent\indent
\begin{figure}
\setlength{\unitlength}{1mm}
\centering
\begin{minipage}{0.23\textwidth}
\includegraphics[width=1.5in]{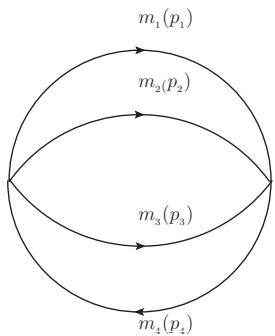}
\end{minipage}%
\vspace{-0.0cm}
\hspace{0cm}
\caption[]{three-loop vacuum diagram, $m_{_i}$ denotes the mass of the $i$-th particle
and $p_{_j}$ denotes the corresponding  momentum.}
\label{c7}
\end{figure}
For scalar integrals, we can get the analytic expressions through the hypergeometric theory.
According to the series representations of modified Bessel functions
and some integrals in hypergeometric theory, our previous work\cite{Feng2017}
have obtained the generalized hypergeometric functions of
the one-loop $B_{_0}$ function, two-loop vacuum integral,
the scalar integrals from two-loop sunset and one-loop triangle diagrams.
And we establish the systems of linear homogeneous PDEs satisfied by the
scalar integrals in the kinematic region.
Furthermore, the $C_{0}$ function have been calculated\cite{Feng2018} under the guidance of MB representations.
The continuation to the whole kinematic domain can be finished with help of finite element.
The point specified here is that the system of homogeneous linear PDEs differs
from that presented in literatures \cite{A.V.Kotikov91-1,A.V.Kotikov91-2,
A.V.Kotikov91-3,A.V.Kotikov91-4,A.V.Kotikov92-1,S.Laporta96,S.Laporta97,
E.Remiddi97,S.Laporta00,K.Melnikov00-1,K.Melnikov00-2}. The detailed description is in Ref.\cite{Feng2017}.
Note that the three-loop vacuum integrals with arbitrary masses are considered numerically in Refs.\cite{Martin2017, A.F.2016},
which are the different methods compared with ours. The results of this paper are consistent with the corresponding results of Ref.\cite{Martin2017}.

\indent\indent
This paper aims at computing the scalar integral of three-loop vacuum  diagram FIG.1.
Our presentation is organized as follows.  In scetion \ref{sec2},
The equivalence of traditional Feynman parametrization and
the hypergeometric theory for this scalar integral is proved.
In scetion \ref{sec3}, we obtain the generalized hypergeometric functions
in independent kinematic variables for the scalar integral, which are convergent
in the connected region. Meanwhile, we write down the
systems of homogeneous linear partial differential equations (PDEs) satisfied by
the corresponding the generalized hypergeometric functions.
According to the PDEs, the continuation can be finished from the convergent domain
to whole kinematic regions with the help of the finite element method.
As the special case, the analytical results of the three-loop vacuum integral
in the convergence region are presented in section \ref{sec4}.
Finally our conclusions are summarized in section \ref{sec5}.
\section{The equivalency between Feynman parametrization
and the hypergeometric method\label{sec2}}
\indent\indent
The modified Bessel functions can be
written in the following form \cite{G.N.Watson44,E1,S1,S2}
\begin{eqnarray}
&&{2(m^2)^{D/2-\alpha}\over(4\pi)^{D/2}\Gamma(\alpha)}
k_{_{D/2-\alpha}}(mx)
=\int{d^Dq\over(2\pi)^D}{\exp[-i{\bf q}\cdot {\bf x}]\over(q^2+m^2)^\alpha}\;,\nonumber\\
&&{\Gamma(D/2-\alpha)\over(4\pi)^{D/2}\Gamma(\alpha)}({x\over2})^{2\alpha-D}
=\int{d^Dq\over(2\pi)^D}{\exp[-i{\bf q}\cdot {\bf x}]\over(q^2)^\alpha}\;,
\label{U1}
\end{eqnarray}
where ${\bf q}$, ${\bf x}$ are vectors in the $D-$dimension Euclid space, $m$ is the mass of the corresponding particle.
After trivial cancellations of numerator and denominator terms, the general analytic expression for
 the scalar integral of three-loop vacuum diagram FIG.1
can be written in the form
\begin{eqnarray}
&&U_{_3}(m_{_1}^2,m_{_2}^2,m_{_3}^2,m_{_4}^2)\nonumber\\
&&=(\mu^2)^{6-3D/2}\int\prod\limits_{i=1}^3{d^Dp_{_i}\over(2\pi)^D}
{1\over(p_{_1}^2-m_{_1}^2)(p_{_2}^2-m_{_2}^2)(p_{_3}^2-m_{_3}^2)((p_{_1}+p_{_2}+p_{_3})^2-m_{_4}^2)},
\label{U2}
\end{eqnarray}
where $D=4-2\varepsilon$ is the number of dimensions in dimensional regularization and $\mu$ denotes the renormalization energy scale.
With the Wick rotation and Eq.(\ref{U1}), the three-loop $U_{_3}$ function is formulated as
\begin{eqnarray}
&&U_{_3}(m_{_1}^2,m_{_2}^2,m_{_3}^2,m_{_4}^2)
={(-i)2^4(\mu^2)^{6-3D/2}\over\Gamma(D/2)(4\pi)^{3D/2}}
\prod\limits_{i=1}^4(m_{_i}^2)^{D/2-1}\int_0^\infty dx({x\over2})^{D-1}k_{_{D/2-1}}(m_{_i}x)\;.
\label{U3}
\end{eqnarray}
The integral representation of the Bessel function can be applied to Eq.(\ref{U3})
\begin{eqnarray}
&&k_{_\mu}(x)={1\over2}\int_0^\infty t^{-\mu-1}\exp\{-t-{x^2\over4t}\}dt\;,
\;\;\;\Re(x^2)>0\;.
\label{U4}
\end{eqnarray}
Thus the $U_{_3}$ function is written as
\begin{eqnarray}
&&U_{_3}=
{(-i)(\sum\limits_{i=1}^4m_{_i}^2)^{D/2-1}2^{1-D}(\mu^2)^{3D/2-6}\over(4\pi)^{3D/2}\Gamma(D/2)}\int_0^\infty dt_{_1}t_{_1}^{-D/2}
\int_0^\infty dt_{_2}t_{_2}^{-D/2}\int_0^\infty dt_{_3}t_{_3}^{-D/2}\int_0^\infty dt_{_4}t_{_4}^{-D/2}
\nonumber\\
&&\hspace{1.0cm}\times
\exp\{-t_{_1}-t_{_2}-t_{_3}-t_{_4}\}\int_0^\infty dxx^{D-1}
\exp\{(-{m_{_1}^2\over4t_{_1}}-{m_{_2}^2\over4t_{_2}}-{m_{_3}^2\over4t_{_3}}-{m_{_4}^2\over4t_{_4}})x^2\}
\nonumber\\
&&\hspace{0.5cm}={(-i)(\mu^2)^{6-3D/2}\over(4\pi)^{2D}}\int_0^\infty dt_{_1}t_{_1}^{-D/2}
\int_0^\infty dt_{_2}t_{_2}^{-D/2}\int_0^\infty dt_{_3}t_{_3}^{-D/2}\int_0^\infty dt_{_4}t_{_4}^{-D/2}
\nonumber\\
&&\hspace{1.0cm}\times
\exp\{-m_{_1}^2t_{_1}-m_{_2}^2t_{_2}-m_{_3}^2t_{_3}-m_{_4}^2t_{_4}\}
\int d^Dx\exp\{-{t_{_1}t_{_2}t_{_3}+t_{_1}t_{_2}t_{_4}+t_{_1}t_{_3}t_{_4}+t_{_2}t_{_3}t_{_4}\over4t_{_1}t_{_2}t_{_3}t_{_4}}x^2\}
\nonumber\\
&&\hspace{0.5cm}={(-i)(\mu^2)^{6-3D/2}\over(4\pi)^{3D/2}}\int_0^\infty dt_{_1}t_{_1}^{-D/2}
\int_0^\infty dt_{_2}t_{_2}^{-D/2}\int_0^\infty dt_{_3}t_{_3}^{-D/2}\int_0^\infty dt_{_4}t_{_4}^{-D/2}
\nonumber\\
&&\hspace{1.0cm}\times
\exp\{-m_{_1}^2t_{_1}-m_{_2}^2t_{_2}-m_{_3}^2t_{_3}-m_{_4}^2t_{_4}\}
({t_{_1}t_{_2}t_{_3}t_{_4}\over t_{_1}t_{_2}t_{_3}+t_{_1}t_{_2}t_{_4}+t_{_1}t_{_3}t_{_4}+t_{_2}t_{_3}t_{_4}})^{D/2}\;.
\label{U5}
\end{eqnarray}
Performing the variable transformation
\begin{eqnarray}
&&t_{_1}=\varrho y_{_1},\;t_{_2}=\varrho y_{_2},\;t_{_3}=\varrho y_{_3},\;t_{_4}=\varrho (1-y_{_1}-y_{_2}-y_{_3}),\;
\label{U6}
\end{eqnarray}
the Jacobian of the transformation is
\begin{eqnarray}
&&{\partial(t_{_1},t_{_2},t_{_3},t_{_4})\over\partial(y_{_1},y_{_2},y_{_3},\varrho)}=\varrho^3\;,
\label{U7}
\end{eqnarray}
finally we have
\begin{eqnarray}
&&U_{_3}={(-i)(\mu^2)^{6-3D/2}\over(4\pi)^{3D/2}}\int_0^1dy_{_1}\int_0^1dy_{_2}\int_0^1dy_{_3}\nonumber\\
&&\hspace{1.0cm}\times
\int_0^\infty d\varrho^{3-3D/2}\exp\{(-m_{_1}^2y_{_1}-m_{_2}^2y_{_2}-m_{_3}^2y_{_3}-m_{_4}^2(1-y_{_1}-y_{_2}-y_{_3}))\varrho\}
\nonumber\\
&&\hspace{1.0cm}\times
({1\over y_{_1}y_{_2}y_{_3}+y_{_1}y_{_2}(1-y_{_1}-y_{_2}-y_{_3})
+y_{_1}y_{_3}(1-y_{_1}-y_{_2}-y_{_3})+y_{_2}y_{_3}(1-y_{_1}-y_{_2}-y_{_3})})^{D/2}
\nonumber\\
&&\hspace{0.6cm}=
{(-i)\Gamma(4-3D/2)\over(4\pi)^{3D/2}(\mu^2)^{3D/2-6}}\int_0^1dy_{_1}\int_0^1dy_{_2}\int_0^1dy_{_3}\int_0^1dy_{_4}\delta(1-y_{_1}-y_{_2}-y_{_3}-y_{_4})
\nonumber\\
&&\hspace{1.0cm}\times
{(m_{_1}^2y_{_1}+m_{_2}^2y_{_2}+m_{_3}^2y_{_3}+m_{_4}^2y_{_4})^{3D/2-4}\over(y_{_1}y_{_2}y_{_3}+y_{_1}y_{_2}y_{_4}
+y_{_1}y_{_3}y_{_4}+y_{_2}y_{_3}y_{_4})^{D/2}}\;.
\label{U3hyp}
\end{eqnarray}
One can get the same result as Eq.(\ref{U3hyp}) while using the Feynman parametrization for $U_{_3}$ function.

\section{The system of homogeneous linear PDEs for three-loop vacuum diagram FIG.1: general case \label{sec3}}
In order to obtain the triple hypergeometric series for the scalar integral from three-loop vacuum diagram,
the modified Bessel functions in power series an be written as
\begin{eqnarray}
&&k_{_{\mu}}(x)={1\over2}\sum\limits_{n=0}^\infty{(-1)^n\over n!}\Big[
\Gamma(-\mu-n)\Big({x\over2}\Big)^{2n}+\Gamma(\mu-n)\Big({x\over2}\Big)^{2(n-\mu)}\Big]
\nonumber\\
&&\hspace{1.1cm}=
{\Gamma(\mu)\Gamma(1-\mu)\over2}\sum\limits_{n=0}^\infty{1\over n!}\Big[
-{1\over\Gamma(1+\mu+n)}\Big({x\over2}\Big)^{2n}+{1\over\Gamma(1-\mu+n)}\Big({x\over2}\Big)^{2(n-\mu)}\Big]\;.
\label{U8}
\end{eqnarray}
And one can present radial integral\cite{G.N.Watson44} as,
\begin{eqnarray}
&&\int_0^\infty dt \Big({t\over2}\Big)^{2\varrho-1}k_{_\mu}(t)={1\over2}
\Gamma(\varrho)\Gamma(\varrho-\mu)\;.
\label{U9}
\end{eqnarray}
According to the topology of FIG.1, we can get the similar analytic expression whichever $m_{_i}(i =
1,2,3,4)$ is the maximum mass. So, we take the $m_{_4}$ maximum mass as an example to illustrate the calculation process.
As $m_{_4}>max(m_{_1},m_{_2},m_{_3})$, inserting the expressions of
$k_{_{D/2-1}}(m_{_1}x)$, $k_{_{D/2-1}}(m_{_2}x)$,
$k_{_{D/2-1}}(m_{_3}x)$  into the Eq.(\ref{U3}) and then
applying Eq.(\ref{U9}), the $U_{_3}$ function is written as
\begin{eqnarray}
&&U_{_3}(m_{_1}^2,m_{_2}^2,m_{_3}^2,m_{_4}^2)\nonumber\\
&&=-{i({m_{_4}^2})^{3D/2-4}(\mu^2)^{6-3D/2}\over\Gamma(D/2)(4\pi)^{3D/2}}\Gamma^3({D\over2}-1)\Gamma^3(2-{D\over2})
\phi(x_{_1},x_{_2},x_{_3})\;,
\label{U10}
\end{eqnarray}
with $x_{_1}={m_{_1}^2\over m_{_4}^2}$, $x_{_2}={m_{_2}^2\over m_{_4}^2}$, $x_{_3}={m_{_3}^2\over m_{_4}^2}$,
and the function $\phi(x_{_1},x_{_2},x_{_3})$ is defined as
\begin{eqnarray}
&&\phi(x_{_1},x_{_2},x_{_3})=-{1\over\Gamma^2(D/2)}(x_{_1}x_{_2}x_{_3})^{D/2-1}
F_{_C}^{(3)}\left(\left.\begin{array}{ccc}1,&D/2;&\;\\
D/2,&D/2,&D/2;\end{array}\right|x_{_1},x_{_2},x_{_3}\right)
\nonumber\\
&&\hspace{2.5cm}
+{1\over\Gamma^2(D/2)}(x_{_1}x_{_2})^{D/2-1}
F_{_C}^{(3)}\left(\left.\begin{array}{ccc}1,&2-D/2;&\;\\
D/2,&D/2,&2-D/2;\end{array}\right|x_{_1},x_{_2},x_{_3}\right)
\nonumber\\
&&\hspace{2.5cm}
+{1\over\Gamma^2(D/2)}(x_{_1}x_{_3})^{D/2-1}
F_{_C}^{(3)}\left(\left.\begin{array}{ccc}1,&2-D/2;&\;\\
D/2,&2-D/2,&D/2;\end{array}\right|x_{_1},x_{_2},x_{_3}\right)
\nonumber\\
&&\hspace{2.5cm}
+{1\over\Gamma^2(D/2)}(x_{_2}x_{_3})^{D/2-1}
F_{_C}^{(3)}\left(\left.\begin{array}{ccc}1,&2-D/2;&\;\\
2-D/2,&D/2,&D/2;\end{array}\right|x_{_1},x_{_2},x_{_3}\right)
\nonumber\\
&&\hspace{2.5cm}
-{\Gamma(3-D)\over\Gamma(D/2)\Gamma(2-D/2)}(x_{_1})^{D/2-1}
F_{_C}^{(3)}\left(\left.\begin{array}{ccc}2-D/2,&3-D;&\;\\
D/2,&2-D/2,&2-D/2;\end{array}\right|x_{_1},x_{_2},x_{_3}\right)
\nonumber\\
&&\hspace{2.5cm}
-{\Gamma(3-D)\over\Gamma(D/2)\Gamma(2-D/2)}(x_{_2})^{D/2-1}
F_{_C}^{(3)}\left(\left.\begin{array}{ccc}2-D/2,&3-D;&\;\\
2-D/2,&D/2,&2-D/2;\end{array}\right|x_{_1},x_{_2},x_{_3}\right)
\nonumber\\
&&\hspace{2.5cm}
-{\Gamma(3-D)\over\Gamma(D/2)\Gamma(2-D/2)}(x_{_3})^{D/2-1}
F_{_C}^{(3)}\left(\left.\begin{array}{ccc}2-D/2,&3-D;&\;\\
2-D/2,&2-D/2,&D/2;\end{array}\right|x_{_1},x_{_2},x_{_3}\right)
\nonumber\\
&&\hspace{2.5cm}
+{\Gamma(3-D)\Gamma(4-3D/2)\over\Gamma^3(2-D/2)}
F_{_C}^{(3)}\left(\left.\begin{array}{ccc}3-D,&4-3D/2;&\;\\
2-D/2,&2-D/2,&2-D/2;\end{array}\right|x_{_1},x_{_2},x_{_3}\right)\;.
\label{U11}
\end{eqnarray}
Here $F_{_C}^{(3)}$ is the Lauricella function of three independent variables
\begin{eqnarray}
&&F_{_C}^{(3)}\left(\left.\begin{array}{ccc}a,&b;&\;\\
c_{_1},&c_{_2},&c_{_3};\end{array}\right|x_{_1},x_{_2},x_{_3}\right)
=\sum\limits_{n_{_1}=0}^\infty\sum\limits_{n_{_2}=0}^\infty
\sum\limits_{n_{_3}=0}^\infty{(a)_{_{n_{_1}+n_{_2}+n_{_3}}}(b)_{_{n_{_1}+n_{_2}+n_{_3}}}
\over n_{_1}!n_{_2}!n_{_3}!(c_{_1})_{_{n_{_1}}}(c_{_2})_{_{n_{_2}}}(c_{_3})_{_{n_{_3}}}}
x_{_1}^{n_{_1}}x_{_2}^{n_{_2}}x_{_3}^{n_{_3}}\;,
\hspace{1cm}
\label{Lauricella}
\end{eqnarray}
with the connected convergent region $\sqrt{|x_{_1}|}+\sqrt{|x_{_2}|}+\sqrt{|x_{_3}|}\leq1$.

For the case $m_{_3}>max(m_{_1},m_{_2},m_{_4})$, one similarly derives
\begin{eqnarray}
&&U_{_3}(m_{_1}^2,m_{_2}^2,m_{_3}^2,m_{_4}^2)\nonumber\\
&&=-{i({m_{_3}^2})^{3D/2-4}(\mu^2)^{6-3D/2}\over\Gamma(D/2)(4\pi)^{3D/2}}\Gamma^3({D\over2}-1)\Gamma^3(2-{D\over2})
\phi(y_{_1},y_{_2},y_{_3})\;,
\label{U12}
\end{eqnarray}
with $y_{_1}={m_{_1}^2\over m_{_3}^2}={x_{_1}\over x_{_3}}$, $y_{_2}={m_{_2}^2\over m_{_3}^2}={x_{_2}\over x_{_3}}$,
$y_{_3}={m_{_4}^2\over m_{_3}^2}={1\over x_{_3}}$.
We specify here that $\phi(y_{_1},y_{_2},y_{_3})=(x_{_3})^{4-3D/2}\phi(x_{_1},x_{_2},x_{_3})$.
For the case $m_{_2}>max(m_{_1},m_{_3},m_{_4})$ and $m_{_1}>max(m_{_2},m_{_3},m_{_4})$ , we can get the similar results.

In other words, when $D=4-2\varepsilon$, the analytic expression of the $U_{_3}$ function
can be formulated as
\begin{eqnarray}
&&U_{_3}(m_{_1}^2,m_{_2}^2,m_{_3}^2,m_{_4}^2)\nonumber\\
&&=-{i\Gamma^3(1-\varepsilon)\Gamma^3(\varepsilon)\over\Gamma(2-\varepsilon)(4\pi)^4}
\Big({m_{_4}^2\over4\pi}\Big)^{2-3\varepsilon}(\mu^2)^{3\varepsilon}\Phi_{_U}(x_{_1},x_{_2},x_{_3})\;,
\label{U13}
\end{eqnarray}
where
\begin{eqnarray}
&&\Phi_{_U}(x_{_1},x_{_2},x_{_3})=\left\{\begin{array}{ll}
\phi(x_{_1},x_{_2},x_{_3})\;,&\sqrt{|x_{_1}|}+\sqrt{|x_{_2}|}+\sqrt{|x_{_3}|}\leq1\;,\\
(x_{_3})^{3D/2-4}\phi({x_{_1}\over x_{_3}},{x_{_2}\over x_{_3}},{1\over x_{_3}})\;,&1+\sqrt{|x_{_1}|}+\sqrt{|x_{_2}|}\leq\sqrt{|x_{_3}|}\;,\\
(x_{_2})^{3D/2-4}\phi({x_{_1}\over x_{_2}},{x_{_3}\over x_{_2}},{1\over x_{_2}}),&
1+\sqrt{|x_{_1}|}+\sqrt{|x_{_3}|}\leq\sqrt{|x_{_2}|}\;,\\
(x_{_1})^{3D/2-4}\phi({x_{_3}\over x_{_1}},{x_{_2}\over x_{_1}},{1\over x_{_1}}),&
1+\sqrt{|x_{_2}|}+\sqrt{|x_{_3}|}\leq\sqrt{|x_{_1}|}\;.\\
\end{array}\right.
\label{sysfunction1}
\end{eqnarray}
Here, $\Phi_{_U}(x_{_1},x_{_2},x_{_3})$ satisfies the system of homogeneous linear PDEs\cite{France}
\begin{eqnarray}
&&\Big[(\sum_{i=1}^3\hat{\vartheta}_{x_{_i}}+3-D)(\sum_{i=1}^3\hat{\vartheta}_{x_{_i}}+4-{3D\over 2})
-{1\over x_{_i}}\hat{\vartheta}_{x_{_i}}(\hat{\vartheta}_{x_{_i}}+1-{D\over2})\Big]\Phi_{_U}(x_{_1},x_{_2},x_{_3})=0\:,
\label{1PDES}
\end{eqnarray}
with $\hat{\vartheta}_{x_{_i}}=x_{_i}{\partial/\partial x_{_i}}\;,i=1,2,3.$\\
 \indent\indent
The $\Phi_{_U}$ function under the restriction $x_{_2}=x_{_3}=0$
is given as
\begin{eqnarray}
&&\Phi_{_U}(x_{_1},0,0)=F(x_{_1})=\left\{\begin{array}{ll}
\phi(x_{_1},0,0)\;,&|x_{_1}|\leq1\\
(x_{_1})^{3D/2-4}\phi({1\over x_{_1}},0,0)\;,&|x_{_1}|\geq1
\end{array}\right.\;.
\label{U14}
\end{eqnarray}
And one derives $\phi(x_{_1},0,0)=(x_{_1})^{3D/2-4}\phi({1\over x_{_1}},0,0)$.
In the whole $x_{_1}-$coordinate axis, $F(x_{_1})$ is a continuous differentiable function.
And $F(x_{_1})$ satisfies the first PDE with the condition $x_{_2}=x_{_3}=0$ in Eq.(\ref{1PDES}).
In the similarly way, one can get the analytic expressions for $F(x_{_2}), F(x_{_3})$
in the whole $x_{_2}, x_{_3}-$coordinate axis.
Taking the $\Phi_{_U}(x_{_1},0,0)=F(x_{_1})$ and $\Phi_{_U}(0,x_{_2},0)=F(x_{_2})$ as boundary conditions,
we can get the numerical solutions of $\Phi_{_U}$ on the entire $x_{_1}-x_{_2}$ plane by the first two PDEs
when $x_{_3}=0$.
Using the similar method, the continuation of $\Phi_{_U}$ to whole three dimension space
can be finished through the system of PDEs in Eq.(\ref{1PDES}).

\indent\indent
We give the Laurent series of $\Phi_{_U}$ function around space-time dimensions $D=4$
in order to make the numerical continuation of $\Phi_{_U}$ to whole kinematic regions,
\begin{eqnarray}
&&\Phi_{_U}(x_{_1},x_{_2},x_{_3})={\phi_{_U}^{(-3)}(x_{_1},x_{_2},x_{_3})\over\varepsilon^3}+
{\phi_{_U}^{(-2)}(x_{_1},x_{_2},x_{_3})\over\varepsilon^2}
+{\phi_{_U}^{(-1)}(x_{_1},x_{_2},x_{_3})\over\varepsilon}
+\sum\limits_{i=0}^\infty\varepsilon^i\phi_{_U}^{(i)}(x_{_1},x_{_2},x_{_3})\;.\nonumber\\
\label{U15}
\end{eqnarray}
The systems of linear PDEs in appendix \ref{app1}
are derived, which satisfied by the functions $\phi_{_U}^{(-3)}$, $\phi_{_U}^{(-2)}$
and $\phi_{_U}^{(i)}(i=-1,\;0,\;1,\;2,\;\cdots)$.\\
\indent\indent
Through the systems of PDEs in appendix \ref{app1}, the continuation of the numerical solution of the triple hypergeometric series
can be made to whole kinematic domain. One derives
$\phi_{_U}^{(-3)}=(x_{_1}+x_{_2}+x_{_3})/2$ satisfies Eq.(\ref{system-1}) explicitly. After obtaining the solutions $\phi_{_U}^{(n-2)}$ and $\phi_{_U}^{(n-1)}$, one writes $F=x_{_1}^{-1/2}x_{_2}^{-1/2}x_{_3}^{-1/2}\phi_{_U}^{(n)}$ satisfies the system of linear PDEs
\begin{eqnarray}
&&2x_{_1}{\partial^2F\over\partial x_{_1}^2}-x_{_2}{\partial^2F\over\partial x_{_2}^2}
-x_{_3}{\partial^2F\over\partial x_{_3}^2}
+2{\partial F\over\partial x_{_1}}-{\partial F\over\partial x_{_2}}
-{\partial F\over\partial x_{_3}}
+(-{1\over 2x_{_1}}+{1\over4x_{_2}}+{1\over4x_{_3}})F
\nonumber\\
&&\hspace{0.0cm}
-x_{_1}^{-1/2}x_{_2}^{-1/2}x_{_3}^{-1/2}(2g_{_1}-g_{_2}-g_{_3})=0
\;,\nonumber\\
&&x_{_2}{\partial^2F\over\partial x_{_2}^2}
-x_{_3}{\partial^2F\over\partial x_{_3}^2}
+{\partial F\over\partial x_{_2}}-{\partial F\over\partial x_{_3}}
+(-{1\over4x_{_2}}+{1\over4x_{_3}})F
-x_{_1}^{-1/2}x_{_2}^{-1/2}x_{_3}^{-1/2}(g_{_2}-g_{_3})=0
\;,\nonumber\\
&&x_{_1}^2{\partial^2F\over\partial x_{_1}^2}
+x_{_2}^2{\partial^2F\over\partial x_{_2}^2}
+x_{_3}(x_{_3}-1){\partial^2F\over\partial x_{_3}^2}
+2x_{_1}x_{_2}{\partial^2F\over\partial x_{_1}\partial x_{_2}}+2x_{_1}x_{_3}{\partial^2F\over\partial x_{_1}\partial x_{_3}}
+2x_{_2}x_{_3}{\partial^2F\over\partial x_{_2}\partial x_{_3}}
\nonumber\\
&&\hspace{0.0cm}
+x_{_1}{\partial F\over\partial x_{_1}}+x_{_2}{\partial F\over\partial x_{_2}}
+(x_{_3}-1){\partial F\over\partial x_{_3}}+({1\over 4x_{_3}}-{1\over 4})F
-x_{_1}^{-1/2}x_{_2}^{-1/2}x_{_3}^{-1/2}g_{_3}=0\;,
\label{U15-1}
\end{eqnarray}
and
\begin{eqnarray}
&&g_{_1}(x_{_1},x_{_2},x_{_3})=-(1-5x_{_1}){\partial\phi_{_U}^{(n-1)}\over\partial x_{_1}}
+5x_{_2}{\partial\phi_{_U}^{(n-1)}\over\partial x_{_2}}
+5x_{_3}{\partial\phi_{_U}^{(n-1)}\over\partial x_{_3}}
-7\phi_{_U}^{(n-1)}+6\phi_{_U}^{(n-2)}
\;,\nonumber\\
&&g_{_2}(x_{_1},x_{_2},x_{_3})=5x_{_1}{\partial\phi_{_U}^{(n-1)}\over\partial x_{_1}}
-(1-5x_{_2}){\partial\phi_{_U}^{(n-1)}\over\partial x_{_2}}
+5x_{_3}{\partial\phi_{U}^{(n-1)}\over\partial x_{_3}}
-7\phi_{_U}^{(n-1)}+6\phi_{_U}^{(n-2)}
\;,\nonumber\\
&&g_{_3}(x_{_1},x_{_2},x_{_3})=5x_{_1}{\partial\phi_{_U}^{(n-1)}\over\partial x_{_1}}
+5x_{_2}{\partial\phi_{_U}^{(n-1)}\over\partial x_{_2}}
-(1-5x_{_3}){\partial\phi_{_U}^{(n-1)}\over\partial x_{_3}}
-7\phi_{_U}^{(n-1)}+6\phi_{_U}^{(n-2)}\;.
\label{U15-2}
\end{eqnarray}
 The second and third PDEs in Eq.(\ref{U15-1}) are recognized as the constraints of the function $F(x_{_1},x_{_2},x_{_3})$.
The system of PDEs in Eq.(\ref{U15-1}) is recognized as the modified functional based on the constraint variational principle\cite{X.C.Wang03},
\begin{eqnarray}
&&\Pi_{_U}^*(F)=\Pi_{_U}(F)
\nonumber\\
&&\hspace{1.8cm}
+\int\limits_{\Omega}\chi_{_{23}}
\Big\{x_{_2}{\partial^2F\over\partial x_{_2}^2}
-x_{_3}{\partial^2F\over\partial x_{_3}^2}
+{\partial F\over\partial x_{_2}}-{\partial F\over\partial x_{_3}}
+(-{1\over4x_{_2}}+{1\over4x_{_3}})F\nonumber\\
&&\hspace{1.8cm}
-x_{_1}^{-1/2}x_{_2}^{-1/2}x_{_3}^{-1/2}(g_{_2}-g_{_3})\Big\}dx_{_1}dx_{_2}dx_{_3}
\nonumber\\
&&\hspace{1.8cm}
+\int\limits_{\Omega}\chi_{_{123}}
\Big\{x_{_1}^2{\partial^2F\over\partial x_{_1}^2}
+x_{_2}^2{\partial^2F\over\partial x_{_2}^2}
+x_{_3}(x_{_3}-1){\partial^2F\over\partial x_{_3}^2}
+2x_{_1}x_{_2}{\partial^2F\over\partial x_{_1}\partial x_{_2}}
\nonumber\\
&&\hspace{1.8cm}
+2x_{_1}x_{_3}{\partial^2F\over\partial x_{_1}\partial x_{_3}}
+2x_{_2}x_{_3}{\partial^2F\over\partial x_{_2}\partial x_{_3}}
+x_{_1}{\partial F\over\partial x_{_1}}+x_{_2}{\partial F\over\partial x_{_2}}
+(x_{_3}-1){\partial F\over\partial x_{_3}}
\nonumber\\
&&\hspace{1.8cm}
+({1\over 4x_{_3}}-{1\over 4})F
-x_{_1}^{-1/2}x_{_2}^{-1/2}x_{_3}^{-1/2}g_{_3}\Big\}dx_{_1}dx_{_2}dx_{_3}\;,
\label{U15-3}
\end{eqnarray}
where $\chi_{_{23}}(x_{_2},x_{_3}),\;\chi_{_{123}}(x_{_1},x_{_2},x_{_3})$
are Lagrange multipliers. $\Omega$ represents the kinematic
domain of numerical solution can be made, and $\Pi_{_U}(F)$ is the functional
of the first PDE in Eq.(\ref{U15-1}):
\begin{eqnarray}
&&\Pi_{_U}(F)=\int\limits_{\Omega}\Big\{
x_{_1}\Big({\partial F\over\partial x_{_1}}\Big)^2
-{x_{_2}\over2}\Big({\partial F\over\partial x_{_2}}\Big)^2
-{x_{_3}\over2}\Big({\partial F\over\partial x_{_3}}\Big)^2
-\Big[-{1\over 4x_{_1}}+{1\over8x_{_2}}
+{1\over8x_{_3}}\Big]F^2
\nonumber\\
&&\hspace{2.0cm}
+x_{_1}^{-1/2}x_{_2}^{-1/2}x_{_3}^{-1/2}
\Big(2g_{_1}-g_{_2}-g_{_3}\Big)F\Big\}dx_{_1}dx_{_2}dx_{_3}\;.
\label{U15-4}
\end{eqnarray}
Firstly, the corresponding function expression of one coordinate axis is taken as the boundary condition, the numerical solution of the planes can be obtained.
Then the whole kinematic region can get the numerical solution by the finite element method \cite{X.C.Wang03}.
In the following, we will discuss
the special case of the three-loop vacuum integral FIG.1.
\section{special case\label{sec4}}
Now, we will deal with the situations different from above.
\subsection{special case one: $m_{_1}=m_{_2}=m\;,m_{_3}\neq0\;,m_{_4}\neq0$\label{4.1}}
When a variable is equal to one in Eq.(\ref{sysfunction1}), the analytic expression of the $U_{_3}$ function
can be simplified. We adopt the following formulae\cite{L.J.Slater66}
\begin{eqnarray}
&&\;_{_2}F_{_1}\left(\left.\begin{array}{cc}a,&b\\ &c\end{array}\right|
1\right)={\Gamma(c)\Gamma(c-a-b)\over\Gamma(c-a)\Gamma(c-b)}\;,
\label{U31}
\end{eqnarray}
if $Rl(c-a-b)>0)$.
With the help of Eq.(\ref{U31}), the $F_{_C}^{(3)}$ function can turn to be
\begin{eqnarray}
&&F_{_C}^{(3)}\left(\left.\begin{array}{ccc}a,&b;&\;\\
c_{_1},&c_{_2},&c_{_3};\end{array}\right|1,x_{_2},x_{_3}\right)\nonumber\\
&&={\Gamma(c)\Gamma(c-a-b)\over\Gamma(c-a)\Gamma(c-b)}\;F_{2;1}^{4;0}\left(\left.\begin{array}{cccc}a,&b,&{1+a-c_{_1}},&{1+b-c_{_1}};\\
{1+a+b-c_{_1}\over2},&{2+a+b-c_{_1}\over2};&c_{_2},&c_{_3};\end{array}\right|{x_{_2}\over4},{x_{_3}\over4}\right)
\hspace{1cm}
\label{Lauricella2}
\end{eqnarray}
and $F_{2;1}^{4;0}$ is the Kamp$\acute{e}$ de F$\acute{e}$riet function\cite{HaroldExton}
\begin{eqnarray}
F_{C;D}^{A;B}\left(\left.\begin{array}{ccccc}&a_{1},&\cdots, &a_{A};\\b_{1},&b_{1}^{'},&\cdots,&b_{B},&b_{B}^{'};\\
&c_{1},&\cdots, &c_{C};\\d_{1},&d_{1}^{'},&\cdots,&d_{D},&d_{D}^{'};
\end{array}\right|x,y\right)
=\sum_{m=0}^\infty\sum_{n=0}^\infty{}{\prod_{j=1}^{A}(a_j)_{m+n}\prod_{j=1}^{B}(b_j)_{m}(b_j)^{'}_{n}\over \prod_{j=1}^{C}(c_j)_{m+n}\prod_{j=1}^{D}(d_j)_{m}(d_j)^{'}_{n} m!n!}x^my^n\;.
\label{U15-8}
\end{eqnarray}

By the Eq.(\ref{Lauricella2}) and after some simplification, we get the results
\begin{eqnarray}
&&U_{_3}(m^2,m^2,m_{_3}^2,m_{_4}^2)={(-i)(m^2)^{3D/2-4}(\mu^2)^{6-3D/2}\over(4\pi)^{3D/2}\Gamma(D/2)}\omega({x\over4},{y\over4})\;,
\end{eqnarray}
and the $\omega(x,y)$ is given by
\begin{eqnarray}
&&\omega({x\over4},{y\over4})={\Gamma(4-3D/2)\Gamma^2(3-D)\Gamma(2-D/2)\Gamma^2(D/2-1)\over \Gamma(6-2D)}\nonumber\\
&&\hspace{1.5cm}\times\;F_{1;1}^{3;0}\left(\left.\begin{array}{ccc}4-3D/2,&3-D,&2-D/2;\\
7/2-D;&2-D/2,&2-D/2;\end{array}\right|{{x\over4},{y\over4}}\right)\nonumber\\
&&\hspace{1.5cm}
+{\Gamma(3-D)\Gamma^2(2-D/2)\Gamma(D/2-1)\Gamma(1-D/2)\over \Gamma(4-D)}(x)^{D/2-1}\nonumber\\
&&\hspace{1.5cm}\times\;F_{1;1}^{3;0}\left(\left.\begin{array}{ccc}3-D,&2-D/2,&1;\\
5/2-D/2;&D/2,&2-D/2;\end{array}\right|{{x\over4},{y\over4}}\right)\nonumber\\
&&\hspace{1.5cm}+{\Gamma(3-D)\Gamma^2(2-D/2)\Gamma(D/2-1)\Gamma(1-D/2)\over \Gamma(4-D)}(y)^{D/2-1}\nonumber\\
&&\hspace{1.5cm}\times\;F_{1;1}^{3;0}\left(\left.\begin{array}{ccc}3-D,&2-D/2,&1;\\
5/2-D/2;&2-D/2,&D/2;\end{array}\right|{{x\over4},{y\over4}}\right)\nonumber\\
&&\hspace{1.5cm}+{\Gamma(D/2)\Gamma(2-D/2)\Gamma^2(1-D/2)\over \Gamma(2)}(xy)^{D/2-1}\nonumber\\
&&\hspace{1.5cm}\times\;F_{1;1}^{3;0}\left(\left.\begin{array}{ccc}2-D/2,&1,&D/2;\\
3/2;&D/2,&D/2;\end{array}\right|{{x\over4},{y\over4}}\right)\;,
\label{U15-7}
\end{eqnarray}
with $x=m_{_3}^2/ m^2,\;y=m_{_4}^2/ m^2$. And the $F_{1;1}^{3;0}$ convergent region is $\sqrt{|{x\over4}|}+\sqrt{|{y\over4}|}\leq1$.
If we use the method of Mellin-Barnes representation on Eq.(\ref{U2}), we can get the same result as Eq.(\ref{U15-7}). And the function $\omega(x,y)=\omega({1\over4},{1\over4})$ when $m_{_1}=m_{_2}=m_{_3}=m_{_4}=m$.
\subsection{special case two: $m_{_1}=0\;,m_{_2}\neq0\;,m_{_3}\neq0\;,m_{_4}\neq0$\label{4.2}}

We discuss the another special case as $m_{_1}=0$. Assuming $m_{_4}>max(m_{_2},m_{_3})$, the $U_{_3}$ function
can be formulated as
\begin{eqnarray}
&&U_{_3}(0,m_{_2}^2,m_{_3}^2,m_{_4}^2)\nonumber\\
&&={(-i)2^3(m_{_2}^2m_{_3}^2m_{_4}^2)^{D/2-1}(\mu^2)^{6-3D/2}\over(D/2-1)(4\pi)^{3D/2}}
\int_0^\infty dx({x\over2})k_{_{D/2-1}}(m_{_2}x)k_{_{D/2-1}}(m_{_3}x)k_{_{D/2-1}}(m_{_4}x)\nonumber\\
&&={(-i)(m_{_4}^2)^{3D/2-4}(\mu^2)^{6-3D/2}\over(D/2-1)(4\pi)^{3D/2}}\Gamma^2(D/2-1)\Gamma^2(2-D/2)\psi(x_{_1},x_{_2})\;,
\label{U16}
\end{eqnarray}
with $x_{_1}=m_{_2}^2/m_{_4}^2,\;x_{_2}=m_{_3}^2/m_{_4}^2$.
Meanwhile the double
hypergeometric series $\psi(x_{_1},x_{_2})$ is
\begin{eqnarray}
&&\psi(x_{_1},x_{_2})={\Gamma(2-D/2)\over\Gamma(D/2)\Gamma(D/2)}(x_{_1}x_{_2})^{D/2-1}
F_{_4}\left(\left.\begin{array}{cc}1,&2-D/2\;\\
D/2,&D/2;\end{array}\right|x_{_1},x_{_2}\right)\nonumber\\
&&\hspace{2.0cm}-{\Gamma(3-D)\over\Gamma(D/2)}x_{_1}^{D/2-1}
F_{_4}\left(\left.\begin{array}{cc}2-D/2,&3-D\;\\
D/2,&2-D/2;\end{array}\right|x_{_1},x_{_2}\right)\nonumber\\
&&\hspace{2.0cm}-{\Gamma(3-D)\over\Gamma(D/2)}x_{_2}^{D/2-1}
F_{_4}\left(\left.\begin{array}{cc}2-D/2,&3-D\;\\
2-D/2,&D/2;\end{array}\right|x_{_1},x_{_2}\right)\nonumber\\
&&\hspace{2.0cm}+{\Gamma(3-D)\Gamma(4-3D/2)\over\Gamma(2-D/2)\Gamma(2-D/2)}
F_{_4}\left(\left.\begin{array}{cc}3-D,&4-3D/2\;\\
2-D/2,&2-D/2;\end{array}\right|x_{_1},x_{_2}\right)\;,
\label{U17}
\end{eqnarray}
one can get the same result by the method of Mellin-Barnes representation.  And $F_{_4}$ is the Apell function
\begin{eqnarray}
&&F_{_4}\left(\left.\begin{array}{cc}a,&b\\
c_{_1},&c_{_2}\end{array}\right|x_{_1},\;x_{_2}\right)
=\sum\limits_{m=0}^\infty\sum\limits_{n=0}^\infty{(a)_{_{m+n}}(b)_{_{m+n}}
\over m!n!(c_{_1})_{_m}(c_{_2})_{_n}}x_{_1}^mx_{_2}^n\;,
\label{U18}
\end{eqnarray}
whose convergent region is $\sqrt{|x_{_1}|}+\sqrt{|x_{_2}|}\leq1$.
For the case $m_{_3}>\max(m_{_2},m_{_4})$, one similarly derives
\begin{eqnarray}
&&U_{_3}(0,m_{_2}^2,m_{_3}^2,m_{_4}^2)\nonumber\\
&&={(-i)(m_{_3}^2)^{3D/2-4}(\mu^2)^{6-3D/2}\over(D/2-1)(4\pi)^{3D/2}}\Gamma^2(D/2-1)\Gamma^2(2-D/2)\psi(y_{_1},y_{_2})\;,
\label{U19}
\end{eqnarray}
with $y_{_1}=m_{_2}^2/m_{_3}^2=x_{_1}/x_{_2},\;y_{_2}=m_{_4}^2/m_{_3}^2=1/x_{_2}$,
and $\psi(y_{_1},y_{_2})=(x_{_2})^{4-3D/2}\psi(x_{_1},x_{_2})$.
In other words, the analytic expression of the $U_{_3}$ function
can be formulated as
\begin{eqnarray}
&&U_{_3}(0,m_{_2}^2,m_{_3}^2,m_{_4}^2)\nonumber\\
&&=-{i\Gamma^2(1-\varepsilon)\Gamma^2(\varepsilon)\over(1-\varepsilon)(4\pi)^4}
\Big({m_{_4}^2\over4\pi}\Big)^{2-3\varepsilon}(\mu^2)^{3\varepsilon}\Psi_{_U}(x_{_1},x_{_2})\;,
\label{U20}
\end{eqnarray}
where
\begin{eqnarray}
&&\Psi_{_U}(x_{_1},x_{_2})=\left\{\begin{array}{ll}
\psi(x_{_1},x_{_2})\;,&\sqrt{|x_{_1}|}+\sqrt{|x_{_2}|}\leq 1\;,\\
(x_{_2})^{3D/2-4}\psi({x_{_1}\over x_{_2}},{1\over x_{_2}})\;,&1+\sqrt{|x_{_1}|}\leq\sqrt{|x_{_2}|}\;,\\
(x_{_1})^{3D/2-4}\psi({x_{_2}\over x_{_1}},{1\over x_{_1}})\;,&1+\sqrt{|x_{_2}|}\leq\sqrt{|x_{_1}|}\;.
\end{array}\right.
\label{U21}
\end{eqnarray}
Correspondingly the double hypergeometric series
$\Psi_{_U}(x_{_1},x_{_2})$ satisfies the system of homogeneous linear PDEs\cite{France}
\begin{eqnarray}
&&\Big\{(\hat{\vartheta}_{x_{_1}}+\hat{\vartheta}_{x_{_2}}+3-D)
(\hat{\vartheta}_{x_{_1}}+\hat{\vartheta}_{x_{_2}}+4-{3D\over2})
-{1\over x_{_i}}\hat{\vartheta}_{x_{_i}}(\hat{\vartheta}_{x_{_i}}+1
-{D\over2})\Big\}\Psi_{_U}=0\;,
\label{2PDES}
\end{eqnarray}
with $\hat{\vartheta}_{x_{_i}}=x_{_i}{\partial/\partial x_{_i}}\;,i=1,2.$\\
\indent\indent
The $\Psi_{_U}$ function under the restriction
$x_{_2}=0$ is
\begin{eqnarray}
&&\Psi_{_U}(x_{_1},0)=G(x_{_1})=\left\{\begin{array}{ll}
\psi(x_{_1},0)\;,&|x_{_1}|\leq1\\
(x_{_1})^{3D/2-4}\psi({1\over x_{_1}},0)\;,&|x_{_1}|\geq1
\end{array}\right.\;.
\label{U22}
\end{eqnarray}
One can get the analytic expressions $G(x_{_1})$ in the whole $x_{_1}-$coordinate axis.
And the $G(x_{_1})$ function satisfies the first PDE with the restriction $x_{_2}=0$ in Eq.(\ref{2PDES}).
Similarly, $\psi(0,x_{_2})=G(x_{_2})$ satisfies the second PDE with the restriction $x_{_1}=0$ in Eq.(\ref{2PDES}) we can write down.
The continuation of $\Psi_{_U}(x_{_1},x_{_2})$ to the entire $x_{_1}-x_{_2}$ plane can be finished numerically
by its analytic expression on the whole $x_{_i}\;(i=1,2)-$axis and the system of PDEs in Eq.(\ref{2PDES}).\nonumber\\
\indent\indent
We give the Laurent series of $\Psi_{_U}$ function around space-time dimensions $D=4$
in order to make the numerical continuation of $\Psi_{_U}$ to whole kinematic regions,
\begin{eqnarray}
&&\Psi_{_U}(x_{_1},x_{_2})={\psi_{_U}^{(-3)}(x_{_1},x_{_2})\over\varepsilon^2}
+{\psi_{_U}^{(-2)}(x_{_1},x_{_2})\over\varepsilon}
+\sum\limits_{i=-1}^\infty\varepsilon^i\psi_{_U}^{(i)}(x_{_1},x_{_2})\;.
\label{U24}
\end{eqnarray}
The systems of linear PDEs in appendix \ref{app2}
are derived, which satisfied by the functions
$\psi_{_U}^{(-3)}$, $\psi_{_U}^{(-2)}$, $\psi_{_U}^{(-1)}$
and $\psi_{_U}^{(n)}\;(n=0,\;2,\;\cdots)$ .
The numerical continuation of the Apell function can be made by the systems of PDEs in appendix \ref{app2}.
One derives $\psi_{_U}^{(-3)}=(x_{_1}+x_{_2})/2$ which satisfies the system of PDEs in
Eq.(\ref{system-4}) explicitly. After obtaining the solutions $\psi_{_U}^{(n-2)},\;\psi_{_U}^{(n-1)}$ in the whole $x_{_1}-x_{_2}$ plane,
one writes the system of linear PDEs satisfied
by $H=x_{_1}^{-1/2}x_{_2}^{-1/2}\psi_{_U}^{(n)}$ as
\begin{eqnarray}
&&-x_{_1}{\partial^2H\over\partial x_{_1}^2}
+x_{_2}{\partial^2H\over\partial x_{_2}^2}
-{\partial H\over\partial x_{_1}}+{\partial H\over\partial x_{_2}}
+({1\over4x_{_1}}-{1\over4x_{_2}})H
+x_{_1}^{-1/2}x_{_2}^{-1/2}(g_{_1}-g_{_2})=0
\;,\nonumber\\
&&x_{_1}(2x_{_1}-1){\partial^2H\over\partial x_{_1}^2}
+x_{_2}(2x_{_2}-1){\partial^2H\over\partial x_{_2}^2}
+4x_{_1}x_{_2}{\partial^2H\over\partial x_{_1}\partial x_{_2}}
+(2x_{_1}-1){\partial H\over\partial x_{_1}}
\nonumber\\
&&\hspace{0.0cm}
+(2x_{_2}-1){\partial H\over\partial x_{_2}}
+({1\over4x_{_1}}+{1\over4x_{_2}})H
+x_{_1}^{-1/2}x_{_2}^{-1/2}(g_{_1}+g_{_2})=0\;,
\label{U24-1}
\end{eqnarray}
and
\begin{eqnarray}
&&g_{_1}(x_{_1},x_{_2})=-(1-5x_{_1}){\partial\psi_{_U}^{(n-1)}\over\partial x_{_1}}
+5x_{_2}{\partial\psi_{_U}^{(n-1)}\over\partial x_{_2}}
-7\psi_{_U}^{(n-1)}+6\psi_{_U}^{(n-2)}
\;,\nonumber\\
&&g_{_2}(x_{_1},x_{_2})=5x_{_1}{\partial\psi_{_U}^{(n-1)}\over\partial x_{_1}}
-(1-5x_{_2}){\partial\psi_{_U}^{(n-1)}\over\partial x_{_2}}
-7\psi_{_U}^{(n-1)}+6\psi_{_U}^{(n-2)}\;.
\label{U24-2}
\end{eqnarray}
Based on the constraint variational principle\cite{X.C.Wang03}, the system of PDEs
in Eq.(\ref{U24-1}) can be treated as the modified functional with stationary conditions
\begin{eqnarray}
&&\Pi_{_U}^*(H)=\Pi_{_U}(H)
\nonumber\\
&&\hspace{2.0cm}
+\int\limits_{\Omega}\chi_{_{12}}
\Big\{x_{_1}(2x_{_1}-1){\partial^2H\over\partial x_{_1}^2}
+x_{_2}(2x_{_2}-1){\partial^2H\over\partial x_{_2}^2}
+4x_{_1}x_{_2}{\partial^2H\over\partial x_{_1}\partial x_{_2}}
+(2x_{_1}-1){\partial H\over\partial x_{_1}}
\nonumber\\
&&\hspace{2.0cm}
+(2x_{_2}-1){\partial H\over\partial x_{_2}}
+({1\over4x_{_1}}+{1\over4x_{_2}})H
+x_{_1}^{-1/2}x_{_2}^{-1/2}(g_{_1}+g_{_2})\Big\}dx_{_1}dx_{_2}\;,
\label{U24-3}
\end{eqnarray}
where $\chi_{_{12}}(x_{_1},x_{_2})$
are Lagrange multipliers, $\Omega$ represents the kinematic
domain where the numerical continuation is made, and $\Pi_{_U}(F)$ is the functional
of the first PDE in Eq.(\ref{U24-1}):
\begin{eqnarray}
&&\Pi_{_U}(H)=\int\limits_{\Omega}\Big\{
{x_{_1}\over2}\Big({\partial H\over\partial x_{_1}}\Big)^2
-{x_{_2}\over2}\Big({\partial H\over\partial x_{_2}}\Big)^2
+({1\over 8x_{_1}}-{1\over8x_{_2}})H^2
+x_{_1}^{-1/2}x_{_2}^{-1/2}(g_{_1}-g_{_2})H\Big\}dx_{_1}dx_{_2}\;.
\nonumber\\
\label{U24-4}
\end{eqnarray}
Because of the boundary conditions $\Psi_{_U}(x_{_1},0)=G(x_{_1})$,
one can perform the continuation of the solution to whole kinematic region numerically
through finite element method\cite{X.C.Wang03} from Eq.(\ref{U24-3}).\\
\indent\indent
The function $\psi(y_{_1},y_{_2})$ will turn to be $\psi(1,y_{_2})$ as $m_{_2}=m_{_3}=m$. In order to get the simper result, we need the reduction formulae
\cite{EEBoos}
\begin{eqnarray}
&&F_{_4}\left(\left.\begin{array}{cc}a,&b\\
c,&d\end{array}\right|1,\;y\right)
={\Gamma(c)\Gamma(c-a-b)\over\Gamma(c-a)\Gamma(c-b)}{_{_4}F_{_3}}\left(\left.\begin{array}{cccc}a,&b,&1+a-c,&1+b-c\\
&d,&(a+b-c+2)/2,&(a+b-c+1)/2\end{array}\right|{y\over4}\right)\;.\nonumber\\
\label{U24-5}
\end{eqnarray}
 Using the above reduction formula Eq.(\ref{U24-5}), we obtain
\begin{eqnarray}
&&U_{_3}(0,m^2,m^2,m_{_4}^2)\nonumber\\
&&={(-i)(m^2)^{3D/2-4}(\mu^2)^{6-3D/2}\over(4\pi)^{3D/2}\Gamma(D/2-1)}f({y_{_2}\over4})\;,
\label{U24-6}
\end{eqnarray}
with $y_{_2}=m_{_4}^2 /m^2$, and $f({y_{_2}\over4})$ is as follows:
\begin{eqnarray}
&&f({y_{_2}\over4})={\Gamma(4-3D/2)\Gamma^2(3-D)\Gamma(2-D/2)\Gamma^2(D/2-1)\over \Gamma(6-2D)}\nonumber\\
&&\hspace{1.3cm}\times_{_3}F_{_2}\left(\left.\begin{array}{ccc}4-3D/2,&3-D,&2-D/2\\
\;&7/2-D,&2-D/2\end{array}\right|{y_{_2}\over4}\right)\nonumber\\
&&\hspace{1.3cm}+{\Gamma(3-D)\Gamma^2(2-D/2)\Gamma(D/2-1)\Gamma(1-D/2)\over \Gamma(4-D)}(y_{_2})^{D/2-1}\nonumber\\
&&\hspace{1.3cm}\times_{_3}F_{_2}\left(\left.\begin{array}{ccc}1,&3-D,&2-D/2,\\
\;&5/2-D/2,&D/2\end{array}\right|{y_{_2}\over4}\right)\;.
\label{U24-7}
\end{eqnarray}
If $x=0$ in Eq.(\ref{U15-7}), then the result is consistent with Eq.(\ref{U24-7}).
And the Eq.(\ref{U24-7}) is in agreement with Eq.(1.28) of Ref.\cite{A.I.Davydychev}

\subsection{special case three: $m_{_1}=m_{_2}=0\;,m_{_3}\neq0\;,m_{_4}\neq0$\label{4.3}}
When $m_{_1}=m_{_2}=0$,  the $U_{_3}$ function
can be formulated as
\begin{eqnarray}
&&U_{_3}(0,0,m_{_3}^2,m_{_4}^2)\nonumber\\
&&={(-i)2^2(m_{_3}^2m_{_4}^2)^{D/2-1}(\mu^2)^{6-3D/2}\Gamma^2(D/2-1)\over\Gamma(D/2)(4\pi)^{3D/2}}
\int_0^\infty dx({x\over2})^{3-D}k_{_{D/2-1}}(m_{_3}x)k_{_{D/2-1}}(m_{_4}x)\nonumber\\
&&={(-i)(m_{_4}^2)^{3D/2-4}(\mu^2)^{6-3D/2}\Gamma^3(D/2-1)\Gamma(2-D/2)\over\Gamma(D/2)(4\pi)^{3D/2}}\nonumber\\
&&\times\Big\{-{\Gamma(2-D/2)\Gamma(3-D)\over\Gamma(D/2)}x^{D/2-1}
\;_{_2}F_{_1}\left(\left.\begin{array}{cc}2-D/2,&3-D\\
\;&D/2\end{array}\right|{x}\right)\nonumber\\
&&+{\Gamma(3-D)\Gamma(4-3D/2)\over\Gamma(2-D/2)}
\;_{_2}F_{_1}\left(\left.\begin{array}{cc}3-D,&4-3D/2\\
\;&2-D/2\end{array}\right|{x}\right)
\Big\},
\label{U25}
\end{eqnarray}
with $x=m_{_3}^2/m_{_4}^2$. At that case, the result from the method of Mellin-Barnes representation
 coincided with the hypergeometric function method.
And $_{_2}F_{_1}$ is the hypergeometry function
\begin{eqnarray}
\;_{_2}F_{_1}\left(\left.\begin{array}{cc}a,&b\\\;&c\end{array}\right|x\right)
=\sum\limits_{n=0}^\infty{(a)_n(b)_n
\over n!(c)_n}x^n\;,
\hspace{1cm}
\label{U26}
\end{eqnarray}
whose convergent region is $|x|\leq1$, $(a)_n={\Gamma(a+n)\over\Gamma(a)}$.
Using the properties of hypergeometric functions, the analytical results can be extended
from $|x|\leq1$ to the domain of $|x|>1$.
\nonumber\\
\indent\indent
When $m_{_1}=m_{_2}=0$ and $m_{_3}=m_{_4}=m$, with the help of the formulae (\ref{U31}), Eq.(\ref{U25})
can be written as
\begin{eqnarray}
&&U_{_3}(0,0,m^2,m^2)=(-i){(m^2)^2\over(4\pi)^6}
[{1\over\varepsilon^3}\varphi_3^{(-3)}+{1\over\varepsilon^2}\varphi_3^{(-2)} + {1\over\varepsilon} \varphi_3^{(-1)}+\varphi_3^{(0)}+{\cal O}(\varepsilon)]\;,
\label{U302}
\end{eqnarray}
where $\varphi_3^{(-3)}\;,\varphi_3^{(-2)}\;,\varphi_3^{(-1)}\;,\varphi_3^{(0)}$ are written as
\begin{eqnarray}
&&\varphi_3^{(-3)}={1\over 3}\;,\nonumber\\
&&\varphi_3^{(-2)}={1\over 6}(7-6\gamma_{_E}+6 ln{4\pi\mu^2\over m^2})\;,\nonumber\\
&&\varphi_3^{(-1)}={1\over 12}(25-42\gamma_{_E}+18\gamma_{_E}^2+\pi^2+42ln{4\pi\mu^2\over m^2}-36\gamma_{_E}ln{4\pi\mu^2\over m^2}+18ln^2{4\pi\mu^2\over m^2})\;,\nonumber\\
&&\varphi_3^{(0)}={1\over24}(-5-150\gamma_{_E}+126\gamma_{_E}^2 -36\gamma_{_E}^3+7\pi^2-6\gamma_{_E}\pi^2+6ln{4\pi\mu^2\over m^2}
(25-42\gamma_{_E}+18\gamma_{_E}^2+\pi^2)\nonumber\\
&&\hspace{1cm}-18ln^2{4\pi\mu^2\over m^2}(-7+6\gamma_{_E})+36ln^3{4\pi\mu^2\over m^2}+56\zeta(3))\;.
\label{U303}
\end{eqnarray}
When $m_{_1}=m_{_2}=m_{_3}=0\,;m_{_4}=m$, the $U_{_3}$ function
can be formulated as
\begin{eqnarray}
U_3(0,0,0,m^2)=(-i){(m^2)^2\over(4\pi)^6}[{1\over\varepsilon^2}\varphi_4^{(-2)} + {1\over\varepsilon} \varphi_4^{(-1)}+\varphi_4^{(0)}+{\cal O}(\varepsilon)]\;,
\label{U304}
\end{eqnarray}
where $\varphi_4^{(-2)}\;,\varphi_4^{(-1)}\;,\varphi_4^{(0)}$ are written as
\begin{eqnarray}
&&\varphi_4^{(-2)}=-{1\over 12}\;,\nonumber\\
&&\varphi_4^{(-1)}={1\over 8}(-5+2\gamma_{_E}-2ln{4\pi\mu^2\over m^2})\;,\nonumber\\
&&\varphi_4^{(0)}={1\over48}(-145+90\gamma_{_E}-18\gamma_{_E}^2-5\pi^2-90ln{4\pi\mu^2\over m^2}
+36\gamma_{_E}ln{4\pi\mu^2\over m^2}-18ln^2{4\pi\mu^2\over m^2})\;.
\label{U305}
\end{eqnarray}
the result of the Eq.(\ref{U303}) and Eq.(\ref{U305}) are  consistent with the result of Ref.\cite{Martin2017}.
\section{Summary\label{sec5}}
\indent\indent
Using the integral representations for modified Bessel  functions,
we verify the equivalency of Feynman parametrization and
the hypergeometric technique to calculate the scalar integrals
of Feynman diagrams in this work.
For the three-loop vacuum integrals, we have presented only diagram FIG.1 to elucidate the technique in detail.
For scalar integrals of diagram FIG.1, the multiple series of representations which
are convergent in certain connected regions of kinematic invariants can be derived.
The systems of linear homogeneous PDEs can be established by the
scalar integrals in whole kinematic domain.
The continuation of the analytic representations of scalar integrals from the convergent regions
to whole kinematic domain through numerical methods can be performed when recognizing the system of linear PDEs as stationary conditions.
For this purpose, the finite element method can be applied.
For the special case of the three-loop vacuum diagram FIG.1, we derives  the analytic
result in the convergence domain.
We will apply this technique to numerically evaluate the scalar
integrals from multi-loop diagrams elsewhere in the near future.

\begin{acknowledgments}
\indent\indent
We are grateful to professor Tai-Fu Feng for guiding this work. The work has been supported partly by the National Natural
Science Foundation of China (NNSFC) with Grant No. 11535002, No. 11705045, and the youth top-notch talent support program of the Hebei Province.
\end{acknowledgments}

\appendix
\section{The system of linear PDEs for
Laurent expansion around $D=4$\label{app1}}
Correspondingly we present the systems of linear PDEs
satisfied by $\phi_{_U}^{(-3)}$, $\phi_{_U}^{(-2)}$
and $\phi_{_U}^{(n)}\;(n=-1,\;0,\;1,\;\;2,\;\cdots)$:
\begin{eqnarray}
&&\Big\{(\sum\limits_{i=1}^3\hat{\vartheta}_{x_{_i}}-1)
(\sum\limits_{i=1}^3\hat{\vartheta}_{x_{_i}}-2)
-{1\over x_{_1}}\hat{\vartheta}_{x_{_1}}(\hat{\vartheta}_{x_{_1}}-1)
\Big\}\phi_{_U}^{(-3)}=0\;,
\nonumber\\
&&\Big\{(\sum\limits_{i=1}^3\hat{\vartheta}_{x_{_i}}-1)
(\sum\limits_{i=1}^3\hat{\vartheta}_{x_{_i}}-2)
-{1\over x_{_2}}\hat{\vartheta}_{x_{_2}}(\hat{\vartheta}_{x_{_2}}-1)
\Big\}\phi_{_U}^{(-3)}=0\;,
\nonumber\\
&&\Big\{(\sum\limits_{i=1}^3\hat{\vartheta}_{x_{_i}}-1)
(\sum\limits_{i=1}^3\hat{\vartheta}_{x_{_i}}-2)
-{1\over x_{_3}}\hat{\vartheta}_{x_{_3}}(\hat{\vartheta}_{x_{_3}}-1)
\Big\}\phi_{_U}^{(-3)}=0\;,
\label{system-1}
\end{eqnarray}

\begin{eqnarray}
&&\Big\{(\sum\limits_{i=1}^3\hat{\vartheta}_{x_{_i}}-1)
(\sum\limits_{i=1}^3\hat{\vartheta}_{x_{_i}}-2)
-{1\over x_{_1}}\hat{\vartheta}_{x_{_1}}(\hat{\vartheta}_{x_{_1}}-1)
\Big\}\phi_{_U}^{(-2)}
\nonumber\\
&&\hspace{0.0cm}
-\Big\{{1\over x_{_1}}\hat{\vartheta}_{x_{_1}}
-5\sum\limits_{i=1}^3\hat{\vartheta}_{x_{_i}}+7\Big\}\phi_{_U}^{(-3)}=0\;,
\nonumber\\
&&\Big\{(\sum\limits_{i=1}^3\hat{\vartheta}_{x_{_i}}-1)
(\sum\limits_{i=1}^3\hat{\vartheta}_{x_{_i}}-2)
-{1\over x_{_2}}\hat{\vartheta}_{x_{_2}}(\hat{\vartheta}_{x_{_2}}-1)
\Big\}\phi_{_U}^{(-2)}
\nonumber\\
&&\hspace{0.0cm}
-\Big\{{1\over x_{_2}}\hat{\vartheta}_{x_{_2}}
-5\sum\limits_{i=1}^3\hat{\vartheta}_{x_{_i}}+7\Big\}\phi_{_U}^{(-3)}=0\;,
\nonumber\\
&&\Big\{(\sum\limits_{i=1}^3\hat{\vartheta}_{x_{_i}}-1)
(\sum\limits_{i=1}^3\hat{\vartheta}_{x_{_i}}-2)
-{1\over x_{_3}}\hat{\vartheta}_{x_{_3}}(\hat{\vartheta}_{x_{_3}}-1)
\Big\}\phi_{_U}^{(-2)}
\nonumber\\
&&\hspace{0.0cm}
-\Big\{{1\over x_{_3}}\hat{\vartheta}_{x_{_3}}
-5\sum\limits_{i=1}^3\hat{\vartheta}_{x_{_i}}+7\Big\}\phi_{_U}^{(-3)}=0\;,
\label{system-2}
\end{eqnarray}
$$\cdots\;\;\cdots\;\;\cdots\;\;\cdots\;,$$
\begin{eqnarray}
&&\Big\{(\sum\limits_{i=1}^3\hat{\vartheta}_{x_{_i}}-1)
(\sum\limits_{i=1}^3\hat{\vartheta}_{x_{_i}}-2)
-{1\over x_{_1}}\hat{\vartheta}_{x_{_1}}(\hat{\vartheta}_{x_{_1}}-1)
\Big\}\phi_{_U}^{(n)}
\nonumber\\
&&\hspace{0.0cm}
-\Big\{{1\over x_{_1}}\hat{\vartheta}_{x_{_1}}
-5\sum\limits_{i=1}^3\hat{\vartheta}_{x_{_i}}+7\Big\}\phi_{_U}^{(n-1)}
+6\phi_{_U}^{(n-2)}=0\;,
\nonumber\\
&&\Big\{(\sum\limits_{i=1}^3\hat{\vartheta}_{x_{_i}}-1)
(\sum\limits_{i=1}^3\hat{\vartheta}_{x_{_i}}-2)
-{1\over x_{_2}}\hat{\vartheta}_{x_{_2}}(\hat{\vartheta}_{x_{_2}}-1)
\Big\}\phi_{_U}^{(n)}
\nonumber\\
&&\hspace{0.0cm}
-\Big\{{1\over x_{_2}}\hat{\vartheta}_{x_{_2}}
-5\sum\limits_{i=1}^3\hat{\vartheta}_{x_{_i}}+7\Big\}\phi_{_U}^{(n-1)}
+6\phi_{_U}^{(n-2)}=0\;,
\nonumber\\
&&\Big\{(\sum\limits_{i=1}^3\hat{\vartheta}_{x_{_i}}-1)
(\sum\limits_{i=1}^3\hat{\vartheta}_{x_{_i}}-2)
-{1\over x_{_3}}\hat{\vartheta}_{x_{_3}}(\hat{\vartheta}_{x_{_3}}-1)
\Big\}\phi_{_U}^{(n)}
\nonumber\\
&&\hspace{0.0cm}
-\Big\{{1\over x_{_3}}\hat{\vartheta}_{x_{_3}}
-5\sum\limits_{i=1}^3\hat{\vartheta}_{x_{_i}}+7\Big\}\phi_{_U}^{(n-1)}
+6\phi_{_U}^{(n-2)}=0\;,
\label{system-3}
\end{eqnarray}
$$\cdots\;\;\cdots\;\;\cdots\;\;\cdots\;.$$

\section{The system of linear PDEs for
Laurent expansion around $D=4$\label{app2}}
Correspondingly we present the systems of linear PDEs
satisfied by $\psi_{_U}^{(-3)}$, $\psi_{_U}^{(-2)}$
and $\psi_{_U}^{(n)}\;(n=-1,\;0,\;1.\;2,\;\cdots)$:
\begin{eqnarray}
&&\Big\{(\sum\limits_{i=1}^2\hat{\vartheta}_{x_{_i}}-1)
(\sum\limits_{i=1}^2\hat{\vartheta}_{x_{_i}}-2)
-{1\over x_{_1}}\hat{\vartheta}_{x_{_1}}(\hat{\vartheta}_{x_{_1}}-1)
\Big\}\psi_{_U}^{(-3)}=0\;,
\nonumber\\
&&\Big\{(\sum\limits_{i=1}^2\hat{\vartheta}_{x_{_i}}-1)
(\sum\limits_{i=1}^2\hat{\vartheta}_{x_{_i}}-2)
-{1\over x_{_2}}\hat{\vartheta}_{x_{_2}}(\hat{\vartheta}_{x_{_2}}-1)
\Big\}\psi_{_U}^{(-3)}=0\;,
\nonumber\\
\label{system-4}
\end{eqnarray}

\begin{eqnarray}
&&\Big\{(\sum\limits_{i=1}^2\hat{\vartheta}_{x_{_i}}-1)
(\sum\limits_{i=1}^2\hat{\vartheta}_{x_{_i}}-2)
-{1\over x_{_1}}\hat{\vartheta}_{x_{_1}}(\hat{\vartheta}_{x_{_1}}-1)
\Big\}\psi_{_U}^{(-2)}
\nonumber\\
&&\hspace{0.0cm}
-\Big\{{1\over x_{_1}}\hat{\vartheta}_{x_{_1}}
-5\sum\limits_{i=1}^2\hat{\vartheta}_{x_{_i}}+7\Big\}\psi_{_U}^{(-3)}=0\;,
\nonumber\\
&&\Big\{(\sum\limits_{i=1}^2\hat{\vartheta}_{x_{_i}}-1)
(\sum\limits_{i=1}^2\hat{\vartheta}_{x_{_i}}-2)
-{1\over x_{_2}}\hat{\vartheta}_{x_{_2}}(\hat{\vartheta}_{x_{_2}}-1)
\Big\}\psi_{_U}^{(-2)}
\nonumber\\
&&\hspace{0.0cm}
-\Big\{{1\over x_{_2}}\hat{\vartheta}_{x_{_2}}
-5\sum\limits_{i=1}^2\hat{\vartheta}_{x_{_i}}+7\Big\}\psi_{_U}^{(-3)}=0\;,
\nonumber\\
\label{system-5}
\end{eqnarray}
$$\cdots\;\;\cdots\;\;\cdots\;\;\cdots\;,$$
\begin{eqnarray}
&&\Big\{(\sum\limits_{i=1}^2\hat{\vartheta}_{x_{_i}}-1)
(\sum\limits_{i=1}^2\hat{\vartheta}_{x_{_i}}-2)
-{1\over x_{_1}}\hat{\vartheta}_{x_{_1}}(\hat{\vartheta}_{x_{_1}}-1)
\Big\}\psi_{_U}^{(n)}
\nonumber\\
&&\hspace{0.0cm}
-\Big\{{1\over x_{_1}}\hat{\vartheta}_{x_{_1}}
-5\sum\limits_{i=1}^2\hat{\vartheta}_{x_{_i}}+7\Big\}\psi_{_U}^{(n-1)}
+6\psi_{_U}^{(n-2)}=0\;,
\nonumber\\
&&\Big\{(\sum\limits_{i=1}^2\hat{\vartheta}_{x_{_i}}-1)
(\sum\limits_{i=1}^2\hat{\vartheta}_{x_{_i}}-2)
-{1\over x_{_2}}\hat{\vartheta}_{x_{_2}}(\hat{\vartheta}_{x_{_2}}-1)
\Big\}\psi_{_U}^{(n)}
\nonumber\\
&&\hspace{0.0cm}
-\Big\{{1\over x_{_2}}\hat{\vartheta}_{x_{_2}}
-5\sum\limits_{i=1}^2\hat{\vartheta}_{x_{_i}}+7\Big\}\psi_{_U}^{(n-1)}
+6\psi_{_U}^{(n-2)}=0\;,
\nonumber\\
\label{system-6}
\end{eqnarray}
$$\cdots\;\;\cdots\;\;\cdots\;\;\cdots\;.$$


\end{document}